\documentstyle[preprint,tighten,aps,graphicx,fancybox]{revtex}
\let\mathbbm\bf
\let\mathscr\mathcal
\title{On (non)linear Quantum Mechanics}
\author{Peter Nattermann\thanks{
  Electronic-mail: {\tt aspn@pt.tu-clausthal.de}}\\[.5ex]
  Institut f\"ur Theoretische Physik\\
  Technische Universit\"at Clausthal\\
  Germany}
\preprint{ASI-TPA/13/97}
\newcommand{\Ref}[1]{(\ref{#1})}
\newcommand{\TE}{\Phi}                              % nonlinear time evolution
\providecommand{\slantfrac}[2]{\kern.1em^{#1}\kern-.3em/\kern-.1em_{#2}}
\newcommand{\quotient}[2]{\raisebox{.5em}{$\ds #1$} \kern-.3em / %\diagup
  \kern-.1em \raisebox{-.5em}{$\ds #2$}}
\newcommand{\rz}{\mathbbm R}
\newcommand{\ie}{\emph{i.e.}}
\newcommand{\eg}{\emph{e.g.}}

\newcommand{\op}[1]{{\mathbf{#1}}}              % operators
\newcommand{\HS}{{\mathscr{H}}}                 % Hilbert space
\newcommand{\Proj}{{\mathit{Proj}}}      % projection operators
\newcommand{\DM}{{\mathcal T}_1^+}              % density matrices
\newcommand{\BS}{{\mathcal B}}                  % Borel field
\newcommand{\MP}{{\mathscr T}}                  % manifold of pure states
\newcommand{\EC}{{\mathcal C}}                  % external conditions
\newcommand{\Ext}{Ext}                                                          % external condition
\newcommand{\PO}{{\mathscr P}}                  % primitive observables
\newcommand{\MG}{{\mathscr M}}                  % motion group
\newcommand{\SO}{{\mathscr E}}                  % set of observables
\renewcommand{\SS}{{\mathcal S}}                        % set of states
\newcommand{\PS}{{\mathcal E}(\SS)}             % set of pure states        
\newcommand{\dt}{\partial_{t}}
\let\hat\widehat
\let\ds\displaystyle
\begin{document}
\maketitle
\begin{abstract}
  We review a possible framework for (non)linear quantum theories, 
  into which linear quantum mechanics fits as well,  and  
  discuss the notion of ``equivalence'' in this setting. Finally, we 
  draw the attention to persisting severe problems of nonlinear 
  quantum theories. 
\end{abstract}
\section{Nonlinearity in quantum mechanics}
Nonlinearity can enter quantum mechanics in various ways, so there are
a number of associations a physicist can have with the term ``nonlinear
quantum mechanics''. Because of this, we shall start with a (certainly
incomplete) list of those ways that we shall not deal with here.

In quantum field theory nonlinearity occurs in the equations of
interacting field operators. These equations may be quantizations of
nonlinear classical field equations (see \eg\ \cite{Segal60}) or
mathematically tractable models as in $\phi^4$-theory. Here, however,
the field operators remain linear, as does the whole quantum
mechanical setup for these quantum field theories.

On a first quantized level, nonlinear terms have been proposed very
early for a phenomenological and semi-classical description of 
self-interactions, \eg\ of electrons in their own electromagnetic
field (see \eg\ \cite{Fermi27}). Being phenomenological these
approaches are build on linear quantum mechanics and use the standard
notion of observables and states. 
For complex systems the linear multi-particle {\scshape Schr\"odinger} 
equation is often replaced by a nonlinear single-particle
{\scshape Schr\"odinger} equation as in the density functional theory of 
solid state physics. 

There have also been attempts to incorporate friction on a
microscopic level using nonlinear {\scshape Schr\"odinger} equations. Many of
these approaches incorporate stochastic frictional forces in the
nonlinear evolution equation for the wavefunctions (see \eg\
\cite{Messer78}). 
% Nonlinear stochastic {\scshape Schr\"odinger} equations have
% also been used to describe open quantum system through stochastic
% processes on the {\scshape Hilbert} space.
 
Contrary to these, we are concerned with a more fundamental role
of nonlinearity in quantum mechanics.  Notable efforts in this direction
have been launched, for example, by {\scshape Bialinycki-Birula} and {\scshape Mycielski}
\cite{BiaMyc76}, {\scshape Weinberg} \cite{Weinbe89a}, and {\scshape 
Doebner} and {\scshape Goldin} \cite{DoeGol92,DoeGol94}.  

\section{Problems of a fundamentally nonlinear nature}
\label{2:sec}
There are evident problems, if we merely replace (naively)
the evolution equation of quantum mechanics, \ie\ the linear {\scshape
  Schr\"odinger} equation, by a nonlinear variant, but stick to the
usual definitions of linear quantum mechanics, like observables being
represented by self-adjoint operators, and states being represented by
density matrices. 

Density matrices $\op{W}\in\DM(\HS)$ represent in general a couple of
different, but \emph{indistinguishable} mixtures of pure states,
\begin{equation}
  \sum_{j} \lambda_j \vert \psi_j\rangle\langle \psi_j\vert
         = \op{W} = 
  \sum_{j} \lambda_j^\prime \vert \psi_j^\prime\rangle\langle 
            \psi_j^\prime\vert \,,\\
\end{equation}
where $\{(\lambda_j,\psi_j)\}_{j=1,\ldots}$ and
$\{(\lambda_j^\prime,\psi_j^\prime)\}_{j=1,\ldots}$ are different
mixtures of pure 
states $\psi_j$ and $\psi_j^\prime$ with weights $\sum_j \lambda_j=1$
and $\sum_j \lambda_j^\prime =1$, respectively. 
This identification of different mixtures is evidently \emph{not}
invariant under a nonlinear time-evolution $\TE_{t}$ of the
wavefunctions,
\begin{equation}
  \sum_{j} \lambda_j \vert \TE_{t}(\psi_j)\rangle\langle 
  \TE_{t}(\psi_j)\vert \stackrel{i.g.}{\neq}   
  \sum_{j} \lambda_j^\prime \vert \TE_{t}(\psi_j^\prime)\rangle\langle 
            \TE_{t}(\psi_j^\prime)\vert \\
\end{equation}
This apparent contradiction has been used by {\scshape Gisin}, {\scshape 
Polchinski}, and others \cite{Gisin90,Polchi91} to predict superluminal 
communications in an EPR-like experiment for \emph{any} nonlinear quantum 
theory. 

Rather than taking this observation as an inconsistency of a nonlinear
quantum theory (\eg\ as in \cite{Weinbe92:book,Gisin95}), 
we take it as an indication, that the notions of
observables and states in a nonlinear quantum theory have to be
adopted appropriately \cite{Luecke95}. If nonlinear quantum mechanics
is to remain a statistical theory, we need a consistent and complete
statistical interpretation of the wave function and the observables, 
and therefore a consistent description of mixed states.

\section{Generalized quantum mechanics}
\label{3:sec}
In view of the intensive studies on nonlinear {\scshape
Schr\"odinger} equations in the last decade it is astonishing to note
that a framework for a consistent framework of nonlinear quantum
theories has already been given by {\scshape Mielnik} in 1974
\cite{Mielni74}. We shall adopt this approach here and develop the
main ingredients of a quantum theory with nonlinear time evolutions of
wavefunctions. 

Our considerations will be based on a fundamental hypothesis on
physical experiments:
\begin{center}
\shadowbox{
\parbox{13cm}{
%\large
   All measurements can in principle be reduced to a change of the 
   dynamics of the system (\eg\ by invoking external fields) 
   and positional measurements.}
}
\end{center}
In fact, this point of view, which  has been taken by
a number of theoretical physicists
\cite{Mielni74,FeyHib65:book,Nelson66,Bell66}, becomes most evident in
scattering experiments, where the localization of particles is
detected (asymptotically) after interaction. 

Based on this hypothesis we build our framework for a nonlinear
quantum theory on three main ``ingredients'' \cite{Natter97a}:

  First, a \emph{topological space}%\footnote{
%   More generally, we could use topological manifolds here 
%   (\emph{manifolds of pure states} \cite{Mielni74}), but for our 
%   purposes it suffices to take linear topological space.} 
  $\MP$ of wavefunctions. In the one
  particle examples below, this topological space is a {\scshape
  Hilbert} space of square integrable functions, but we may also think
  of other function spaces \cite{Natter97b}.

Secondly, the \emph{time evolutions} are given by homeomorphisms of $\MP$,  
  \begin{equation}
    \TE_{t}^{(\Ext)}\colon\, \MP \to \MP \,,
  \end{equation}
  which depend on the time interval $t$ and the external 
  conditions (\eg\ external fields) $\Ext\in\EC$.
  {\scshape Mielnik}'s \emph{motion group} $\MG$ \cite{Mielni74} 
  is the smallest (semi-) group containing \emph{all} time evolutions
  $\TE_{t}^{(\Ext)}$, close in the topology of pointwise convergence. 

Finally, \emph{positional observables} $\PO$ are represented by
  probability measures on physical space $M$, which
  depend on the wavefunction $\phi\in\MP$, \ie\
  $\PO =\bigl\{p_{B}\bigm|B\in\BS(M)\bigr\}$, where  
  \begin{equation}
    p_B : \MP \to [0,1]\,,\qquad \sum_{k=1}^\infty p_{B_k} =
    p_{B}\,,\quad B=\bigcup_{k=1}^\infty B_k,
  \end{equation}
  for disjoint $B_k\in \BS(M)$.

\begin{figure}[t]
\begin{center}
  \includegraphics[angle=270,scale=0.9]{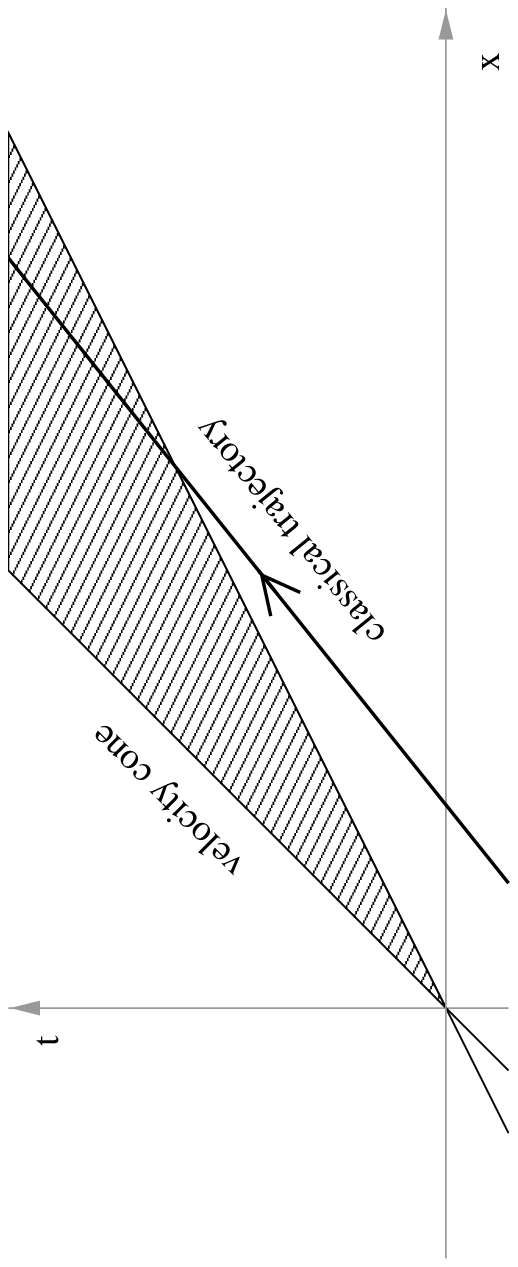}
\end{center}
\caption{Velocity cone for the asymptotic measurement of momentum.}  
\label{fig1}
\end{figure}

We shall call the triple $(\MP,\MG,\PO)$ a \emph{quantum system}. 
Using these basic ingredients we can define effects and
states of the quantum system $(\MP,\MG,\PO)$ as derived
concepts.   
An \emph{effect} (or a \emph{counter}) is (at least approximately in
the sense of pointwise convergence) a 
combination of evolutions $T\in\MG$ and positional measurements $p \in
\PO$, \ie\
\begin{equation}
    \SO := \overline{\left\{ p\circ T \left| p\in\PO,
      T\in\MG\right\}\right. }^{p.c.}
\end{equation}
is the \emph{set of effects}.
A general \emph{observable} $A$ is an $\SO$-valued measure on the set
$M^A$ of its classical values,
\begin{equation}
    p^{A}\colon\, \BS(M^{\! A}) \to \SO\,,\quad p^{A}_{M^{\! A}}[\phi]=1\,.
\end{equation}
The standard example of such an asymptotic observable is the (dynamical)
momentum of a single particle of mass $m$ 
in $\rz^3$. Let $B\in\BS(\rz^3_p)$ be an
open subset of momentum space $\rz_p^3$, then  
\begin{equation}
  B_t := \Bigl\{\frac{t}{m}\vec{p}\,\Bigm\vert\, \vec{p}\in B\Bigr\}\,,
\end{equation}
defines the corresponding velocity cone, see Figure~\ref{fig1}.
If $\TE_{t}^{(0)}$ denotes the free quantum mechanical time evolution
of our theory --- provided, of course, there is such a distinguished
evolution --- the limit 
\begin{equation}
   p^{P}_{B} [\phi] := \lim_{t\to\infty} 
   p_{B_{t}}\Bigl[\TE_{t}^{(0)}[\phi]\Bigr]
\end{equation}
defines a probability measure on $\rz^3_p$, so that the momentum
observable is given by the $\SO$-valued measure  
\begin{equation}
     P = \Bigl\{ p^{P}_{B}\in \SO \Bigm| B\in\BS(\rz^3_p)\Bigr\}\,.
\end{equation}

Coming back to the general framework, once we have determined the set
of effects, we can define the \emph{states} of the quantum system as
equivalence classes of mixtures of wavefunctions.

Different mixtures of wave functions 
\begin{equation}
    \pi = \bigl\{(\lambda_{j},\phi_{j})\bigr\}_{j=1,\ldots}\,,\quad 
    \sum_{j} 
    \lambda_{j} =1\,,
\end{equation}
with corresponding effects $f[\pi] := \sum_{j}\lambda_{j} f(\phi_{j})$
may be \emph{indistinguishable} with respect to the effects $\SO$, 
\begin{equation}
    \pi_{1}\sim \pi_{2}  \quad\Leftrightarrow\quad  \Bigl(f[\pi_{1}] =  
    f[\pi_{2}]\quad\forall\, f\in\SO\Bigr)\,.
\end{equation}
Hence, the \emph{state space}  
\begin{equation}
    \SS := \quotient{\Pi(\MP)}{\sim}
\end{equation}
is a convex set with \emph{pure states} as extremal points $\PS$. 

\section{Linear quantum mechanics}
\label{4:sec}
Generalized quantum mechanics is indeed a generalization of linear
quantum mechanics, as the latter is contained in the general framework
as a special case. To see this, we consider a non-relativistic
particle of mass $m$ in $\rz^3$ ({\scshape Schr\"odinger} particle), 
defined in our setting as a quantum system $(\HS,\MG_{S},\PO_{\chi})$ 
with the {topological space} of wave functions as the {\scshape 
Hilbert}-space  
\begin{equation}
     \MP \equiv \HS \equiv L^2(\rz^3,d^3x)\,,
\end{equation}
the {\scshape Born} interpretation of
$\slantfrac{|\psi(\vec{x})|^2}{\|\psi\|^2}$ as a positional probability 
density on $\rz^3$, \ie\
\begin{equation}
    p_B[\psi] := 
    \frac{\langle\psi\vert\op{E}(B)\psi\rangle}{\|\psi\|^2} =
    \frac{\|\op{E}(B)\psi\|^2}{\|\psi\|^2} 
\end{equation}
defines the positional observables, and unitary time evolutions
generated by linear {\scshape Schr\"odinger} equations
\begin{equation}  
\label{SE:eq}
    i\hbar\dt \psi_t = \left(-\frac{\hbar^2}{2m} \Delta +V\right)\psi_t 
    \equiv \op{H}_{V}\psi_{t}\,,
\end{equation}
with a class of suitable potentials $V$ representing the external conditions of
the system.

Starting with these three objects we recover indeed the full structure
of (linear) quantum mechanics. 
First, the motion group of the {\scshape Schr\"odinger} particle is
the whole unitary group  \cite{Waniew77}
\begin{equation}
    \MG_S \simeq U(\HS)\,.
\end{equation}
Furthermore, the (decision) effects are given precisely by
orthogonal projection operators \cite{Natter97c},
\begin{equation}
  \SO \simeq \Proj(\HS)\,, \quad f_{\op{E}}[\psi] = 
     \frac{\|\op{E}\psi\|^2}{\|\psi\|^2}\,,
\end{equation}
so that the logical structure of quantum mechanics is
recovered; observables occur naturally through their spectral measures
in this scheme.
For example, the asymptotic definition of momentum along the lines given 
above is well known in linear quantum mechanics \cite{Kemble37:book} 
and leads through standard {\scshape Fourier} transform to the usual 
spectral measure of the momentum operator $\op{P}$.

Finally, as a consequence of the above set of effects, 
the \emph{state space} coincides with the space of
normalized, positive trace class operators,
\begin{equation}
     \SS \simeq \DM(\HS) \,,\qquad  \PS \simeq P(\HS)\,.
\end{equation}

\section{Equivalent quantum systems}
Having based our discussion on a fundamental hypothesis on the 
distinguished role of positional measurements in quantum mechanics, 
the notion of \emph{gauge equivalence} has to be reconsidered within 
the generalized framework of the previous section.

As our framework is based on topological spaces, two quantum systems 
$(\MP,\MG,\PO)$ and $(\hat\MP,\hat\MG,\hat\PO)$ are 
\emph{topologically equivalent}, if 
%(i) 
$\PO$ and $\hat\PO$ are positional observables on the same physical space $M$, 
%(ii) 
the time evolutions depend on the same external conditions $\EC$, and 
%(iii) 
there is a homeomorphism $N\colon\, \MP\to \hat\MP$, such that  
\begin{equation}
\begin{array}{rcll}
  p_{B}& = & \hat{p}_{B} \circ N\,,& \forall B 
    \in \BS(M)\,,\\
   {\TE}_{t}^{(\Ext)} & = & N^{-1}\circ 
      \hat{\TE}_{t}^{(\Ext)}\circ N \,,\quad
      &\forall t\in\rz,\,\Ext\in \EC\,.
\end{array}
\end{equation}  
For the linear quantum systems of the previous section this notion of topological 
equivalence reduces naturally to ordinary unitary equivalence.

A particular case arises, if we consider automorphisms of the same 
topological space of wavefunctions $\MP$ that leave the positional 
observables invariant,
\begin{equation}
  N\colon\, \MP\to\MP\,,\qquad p_{B} = p_{B}\circ N \quad \forall B 
    \in \BS(M)\,.
\end{equation} 
We call these automorphisms \emph{generalized gauge transformations}. 
For linear quantum systems, these reduce to ordinary gauge 
transformations of the second kind,
%\begin{equation}
$  \left(\op{U}_{\theta_{t}} \psi_{t}\right)(\vec{x}) = e^{i\theta_{t}(\vec{x})} 
  \psi_{t}(\vec{x})\,.
$
%\end{equation}
As in this linear case, the automorphisms $N$ may be (explicitly) 
time-dependent.  
\section{Quantum mechanics in a nonlinear disguise}
As we have seen in Section~\ref{4:sec}, the framework of 
Section~\ref{3:sec} can in indeed be filled in case of linear evolution 
equations; but are there also nonlinear models?      
{\scshape Mielnik} has listed a number of nonlinear toy models for 
his framework \cite{Mielni74} and has furthermore considered finite 
dimensional nonlinear systems \cite{Mielni80}; {\scshape Haag} and {\scshape 
Bannier} have given an interesting example of a quantum system with 
linear and nonlinear time evolutions \cite{HaaBan78}. 

Here, however, we shall proceed differently in order to obtain nonlinear quantum 
system: We use the generalized gauge transformations introduced in 
the previous section in order to construct nonlinear quantum systems 
$(\HS,\MG,\PO)$ with $L^2$-wavefunctions that are 
\emph{gauge equivalent} to linear quantum mechanics.  

To simplify matters, we assume that the time evolution of the 
nonlinear quantum system is still given by a local, 
(quasi-)homogeneous nonlinear {\scshape Schr\"odinger} equation. 
This leads us to consider \emph{strictly local, projective generalized 
gauge transformations} \cite{Natter97c} 
\begin{equation}
  N_{\gamma_{t}} (\psi_t) = \psi_t 
  \exp\left(i\gamma_t\ln|\psi_t|\right)\,,
\end{equation}
where $\gamma_{t}$ is a time-dependent parameter.
As these automorphisms of $L^2(\rz^3,d^3x)$ are extremely similar to 
local linear gauge transformations of the second kind, they have been 
called \emph{nonlinear gauge transformations} \cite{DoGoNa95a} or gauge 
transformations of the \textrm{third kind} \cite{DoeGol96}.

Using these transformations, the evolution equations for 
$\psi_{t}^\prime := N_{\gamma_{t}} (\psi_t)$, where $\psi_{t}$ is a 
solution of the linear {\scshape Schr\"odinger} equation \Ref{SE:eq} are 
easily calculated:
\begin{equation}
\label{NSE:eq}
\begin{array}{rcl}\ds
  i\hbar \dt \psi_t &=&\ds \left(-\frac{\hbar^2}{2m} \Delta +V\right)\psi_t
     - i\frac{\hbar^2 \gamma_t}{4m} R_{2}[\psi_{t}]\psi_t
     - \frac{\hbar^2 \gamma_t}{4m} \left(R_1[\psi_t]-R_4[\psi_t]\right) \psi_t\\
    &&\ds  
     + \frac{\hbar^2 \gamma_t^2}{16m}
       \left(2R_2[\psi_t]-R_5[\psi_t]\right)\psi_t 
   - \frac{1}{2}\dot\gamma_t \ln |\psi_t|^2\, \psi_t\,,
\end{array}
\end{equation}
where  
\begin{equation}
     R_1[\psi] := \frac{\nabla\cdot \vec{J}}{\rho}\,,\quad
     R_2[\psi] := \frac{\Delta\rho}{\rho}\,,\quad
     R_4[\psi] := \frac{\vec{J}\cdot\nabla\rho}{\rho^2}\,,\quad 
     R_5[\psi] := \frac{\nabla\rho\cdot\nabla\rho}{\rho^2}\,.
\end{equation}
These equations contain typical functionals $R_{j}$ of the {\scshape 
Doebner--Goldin} equations \cite{DoeGol94} as well as the logarithmic term 
of {\scshape Bialynicki-Birula--Mycielski} \cite{BiaMyc76}. Note that 
the form of Eq.~\Ref{NSE:eq} do not immediately reveal its 
linearizability, the underlying linear structure of this model is 
disguised.

In fact, through an iterated process of gauge generalization and gauge 
closure --- similar to the minimal coupling scheme of linear quantum 
mechanics --- we could obtain a unified family of nonlinear {\scshape 
Schr\"odinger} equations \cite{DoeGol96,DoGoNa96} 
($R_3[\psi] := \frac{\vec{J}^{\,2}}{\rho^2}$):
\begin{equation}
\label{USE:eq}
  i\dt \psi_t = i \sum_{j=1}^2 {\nu_j} R_j[\psi_t]\,\psi_t + 
  \mu_{0}V + \sum_{k=1}^5
  {\mu_k} R_k[\psi_t]\, \psi_t + {\alpha_1} \ln |\psi_t|^2\,\psi_t\,.
\end{equation} 

\section{Final Remarks: Histories and Locality}
In this contribution we have sketched a framework for nonlinear 
quantum theories that generalizes the usual linear one. We close 
with three remarks.

The first is concerned with the definition of effects (and positional 
observables) is our framework. Since we have used real-valued 
measures, our observables do not allow for an idealization of measurements as 
in the linear theory, where a projection onto certain parts of the 
spectrum is possible using the projection-valued measure. 
Combined subsequent measurements (\emph{histories}) have to be described by 
quite complicated time evolutions. However, in case of linearizable quantum 
system, \emph{generalized projections} 
\begin{equation}
     E := N \circ \op{E} \circ N^{-1}\,,\qquad \op{E}\in\Proj(\HS)
\end{equation}
onto nonlinear sub-manifolds of $\HS$ can be realized as an 
idealization of measurements, and yield a nonlinear realization of 
the standard quantum logic \cite{Luecke95}.

Secondly, we should emphasize that we have not been able to describe a 
complete and satisfactory nonlinear theory that is \emph{not} 
gauge equivalent to linear quantum mechanics. One of the obstacles of 
quantum mechanical evolution equations like \Ref{USE:eq} is the 
difficulty of the (global) {\scshape Cauchy} problem for partial differential 
equations. Whereas there is a solution for the logarithmic nonlinear {\scshape 
Schr\"odinger} equation \cite{CazHar80}, there are only local 
solutions for (non-linearizable) {\scshape Doebner--Goldin} equations 
\cite{Teisma97a}.

Another problem of nonlinear {\scshape Schr\"odinger} equations in 
quantum mechanics is the locality of the corresponding quantum 
theory: EPR-like experiments could indeed lead to superluminal 
communications, though not in the naive (and irrelevant) fashion 
described in Section~\Ref{2:sec}, relevant {\scshape Gisin}-effects 
\cite{LueNat97} can occur, if %--- because of the nonlinearity --- 
changes of the external conditions in spatially separated regions 
have instantaneous effects.
Since the nonlinear equations we have considered here are 
\emph{separable}, this effect can only occur for \emph{entangled} 
initial wavefunctions, \ie\
\begin{equation}
  V(\vec{x}_1,\vec{x}_2) = V_1(\vec{x}_1) + V_2(\vec{x}_2)\,,\qquad 
  \psi_0(\vec{x}_1,\vec{x}_2) \neq \varphi_1(\vec{x}_1) 
  \varphi_2(\vec{x}_2)\,.  
\end{equation} 
For {\scshape Doebner-Goldin} equations, for instance, such effects 
indeed occur (at least) for certain subfamilies that are not {\scshape 
Galilei} invariant \cite{LueNat97}; (higher order) calculations for the 
{\scshape Galilei} invariant case and the logarithmic {\scshape Schr\"odinger} 
equation are not yet completed.

In the title of this contribution we have put the prefix ``non'' in 
brackets; the remarks above may have indicated why. Finally, one 
might be forced to find \emph{different} ways of extending a nonlinear single 
particle theory to many particles (see \eg\ \cite{Goldin97c,Czacho97a}).

\subsection*{Acknowledgments}
Thanks are due to {\scshape H.-D.~Doebner}, {\scshape G.A.~Goldin}, {\scshape 
W.~L\"ucke}, and {\scshape R.F.~Werner} for interesting and fruitful 
discussions on the subject. I am also greatful to the organizers for 
their invitation to the ``Second International Conference on Symmetry
in Nonlinear Mathematical Physics'' 
It is a pleasure to thank {\scshape Renat Zhdanov}, his family and
friends for their warm hospitality during our stay in Kiev. 


\begin{thebibliography}{10}

\bibitem{Segal60}
Segal, I.~E., {\em J.~Math.~Phys.\,\/}, 1960, V~1, N~6, 468--479.

\bibitem{Fermi27}
Fermi, E., {\em Atti della Reale Accademia nazionale dei Lincei.
  Rendiconti.\/}, 1927, V~5, N~10, 795--800.

\bibitem{Messer78}
Messer, J., {\em Lett.~Math.~Phys.\,\/}, 1978, V~2, 281--286.

\bibitem{BiaMyc76}
Bialynicki-Birula, I. and Mycielski, J., {\em Ann.~Phys.\ (NY)\/}, 1976, V 100,
  62--93.

\bibitem{Weinbe89a}
Weinberg, S., {\em Phys.~Rev.~Lett.\,\/}, 1989, V~62, 485--488.

\bibitem{DoeGol92}
Doebner, H.-D. and Goldin, G.~A., {\em Phys.~Lett.~A\/}, 1992, V 162, 397--401.

\bibitem{DoeGol94}
Doebner, H.-D. and Goldin, G.~A., {\em J.~Phys.~A: Math.~Gen.\/}, 1994, V~27,
  1771--1780.

\bibitem{Gisin90}
Gisin, N., {\em Phys.~Lett.~A\/}, 1990, V 143, 1--2.

\bibitem{Polchi91}
Polchinski, J., {\em Phys.~Rev.~Lett.\,\/}, 1991, V~66, N~4, 397--400.

\bibitem{Weinbe92:book}
Weinberg, S., {\em Dreams of a Final Thoery\/}, Pantheon Books, New York, 1992.

\bibitem{Gisin95}
Gisin, N., Relevant and irrelevant nonlinear {Schrš}dinger equations, in
  Doebner {\em et~al.\/}  \cite{DoDoNa95:proc}, 1995  109--124.

\bibitem{Luecke95}
L{\"u}cke, W., Nonlinear {Schr\"o}dinger dynamics and nonlinear observables, in
  Doebner {\em et~al.\/}  \cite{DoDoNa95:proc}, 1995  140--154.

\bibitem{Mielni74}
Mielnik, B., {\em Comm.~Math.~Phys.\,\/}, 1974, V~37, 221--256.

\bibitem{FeyHib65:book}
Feynman, R. and Hibbs, A., {\em Quantum Mechanics and Path Integrals\/},
  McGraw-Hill Book Company, New York, 1965.

\bibitem{Nelson66}
Nelson, E., {\em Phys.~Rev.\,\/}, 1966, V 150, 1079--1085.

\bibitem{Bell66}
Bell, J.~S., {\em Rev.\ Mod.\ Phys.\,\/}, 1966, V~38, N~3, 447--452.

\bibitem{Natter97a}
Nattermann, P., Generalized quantum mechanics and nonlinear gauge
  transformations, in {\em Symmetry in Science IX\/}, editors B.~Gruber and
  M.~Ramek. Plenum Press, New York, 1997,  269--280, ASI-TPA/4/97,
  quant-ph/9703017.

\bibitem{Natter97b}
Nattermann, P., Nonlinear {Schr\"o}dinger equations via gauge generalization,
  in Doebner {\em et~al.\/}  \cite{DoNaSc97:proc},  428--432, ASI-TPA/9/97.

\bibitem{Waniew77}
Waniewski, J., {\em Rep.~Math.~Phys.\,\/}, 1977, V~11, N~3, 331--337.

\bibitem{Natter97c}
Nattermann, P., {\em Dynamics in {B}orel-Quantization: Nonlinear
  {Schr\"o}dinger Equations vs. Master Equations\/}, Ph.D. thesis, Technische
  Universit{\"a}t Clausthal, June 1997.

\bibitem{Kemble37:book}
Kemble, E.~C., {\em The Fundamental Princeiples of Quantum Mechanics with
  Elementary Applications\/}, Dover Publishing, New York, 1937.

\bibitem{Mielni80}
Mielnik, B., {\em J.~Math.~Phys.\,\/}, 1980, V~21, N~1, 44--54.

\bibitem{HaaBan78}
Haag, R. and Bannier, U., {\em Comm.~Math.~Phys.\,\/}, 1978, V~60, 1--6.

\bibitem{DoGoNa95a}
Doebner, H.-D., Goldin, G.~A., and Nattermann, P., A family of nonlinear
  {Schr\"o}dinger equations: Linearizing transformations and resulting
  structure, in {\em Quantization, Coherent States, and Complex Structures\/},
  editors J.-P. Antoine, S.~Ali, W.~Lisiecki, I.~Mladenov, and A.~Odzijewicz.
  Plenum Publishing Corporation, New York, 1995,  27--31.

\bibitem{DoeGol96}
Doebner, H.-D. and Goldin, G.~A., {\em Phys.~Rev.~A\/}, 1996, V~54, 3764--3771.

\bibitem{DoGoNa96}
Doebner, H.-D., Goldin, G.~A., and Nattermann, P., Gauge transformations in
  quantum mechanics and the unification of nonlinear {Schr\"o}dinger
  equations, ASI-TPA/21/96, quant-ph/9709036, submitted to
  J.~Math.~Phys. 

\bibitem{CazHar80}
Cazenave, T. and Haraux, A., {\em Ann.\ Fac.\ Sc.\ Toul.\/}, 1980, V~11,
  21--51.

\bibitem{Teisma97a}
Teismann, H., The {C}auchy problem for the {D}oebner--{G}oldin equation, in
  Doebner {\em et~al.\/}  \cite{DoNaSc97:proc},  433--438.

\bibitem{LueNat97}
L{\"u}cke, W. and Nattermann, P., Nonlinear quantum mechanics and locality, in
  {\em Symmetry in Science X\/}, editors B.~Gruber and M.~Ramek. Plenum Press,
  New York, 1998, ASI-TPA/12/97, quant-ph/9707055.

\bibitem{Goldin97c}
Goldin, G.~A., Nonlinear gauge transformations and their physical implications,
  talk at Symmetry in ``Nonlinear Mathematical Physics II''.

\bibitem{Czacho97a}
Czachor, M., Complete separability of a class of nonlinear {Schr\"o}dinger and
  {L}iouville-von {N}eumann equations, quant-ph/9708052.

\bibitem{DoDoNa95:proc}
Doebner, H.-D., Dobrev, V.~K., and Nattermann, P., editors, {\em Nonlinear,
  Deformed and Irreversible Quantum Systems\/}. World Scientific, Singapore,
  1995.

\bibitem{DoNaSc97:proc}
Doebner, H.-D., Nattermann, P., and Scherer, W., editors, {\em Group21 ---
  Physical Applications and Mathematical Aspects of Geometry, Groups and
  Algebras\/}, volume~1. World Scientific, Singapore, 1997.

\end{thebibliography}
\end{document}